\documentclass{eas}
\usepackage{graphicx,subfigure,amssymb,caption}
\newcommand\la{\;
  \raise0.3ex\hbox{$<$\kern-0.75em\raise-1.1ex\hbox{$\sim$
  }}\;\hskip-2pt }
\newcommand\ga{\;
  \raise0.3ex\hbox{$>$\kern-0.75em\raise-1.1ex\hbox{$\sim$
  }}\;\hskip-2pt }

\newcommand{\RM}{\mathrm{RM}} 
\newcommand{\DFD}{\Delta F_{D}} 
\newcommand{\DF}{\Delta F} 
\newcommand{\cP}{\mathcal{P}} 
\newcommand{\cF}{\mathcal{F}} 
\newcommand{\er}{\epsilon_{\rho}} 
\newcommand{\esyn}{\epsilon} 
\newcommand{\lam}{\lambda} 
\newcommand{\Lc}{\mathcal{L}} 
\newcommand{\M}{\mathcal{M}} 
\newcommand{\ps}{\psi} 
\newcommand{\pso}{\psi_{0}} 
\newcommand{\po}{p_0} 
\newcommand{\nel}{n_\mathrm{e}} 
\newcommand{\sign}{{\mathrm{sign}}}
\newcommand{\Arctan}{{\mathrm{Arctan}\,}}

\newcommand\sfrac[2]{{\textstyle{\frac{#1}{#2}}}}
\newcommand\mean[1]{\overline{#1}}
\newcommand{\dif}{\mathrm d}
\newcommand\vect[1]{\mathbf{#1}}
\newcommand{\cm}{\,\mathrm{cm}}

\newcommand{\degr}{^{\circ}}

\newcommand{\GHz}{\, \mathrm{GHz}}

\newcommand{\kms}{\,\mathrm{km\,s^{-1}}}

\newcommand{\kpc}{\,\mathrm{kpc}}

\newcommand{\mkG}{\,\mu\mathrm{G}}

\newcommand{\p}{\,\mathrm{pc}}
\newcommand{\radmm}{\, \mathrm{rad}\,\mathrm{m}^{-2}} 

\newcommand{\yr}{\,\mathrm{yr}}

\begin{document}

\title{Depolarization canals and interstellar turbulence}
\runningtitle{A.~Fletcher \& A.~Shukurov: Depolarization canals}
\author{Andrew Fletcher}
\address{School of Mathematics and Statistics,
University of Newcastle, Newcastle upon Tyne, NE1 7RU, U.K;
\email{andrew.fletcher@ncl.ac.uk\ {\rm{(AF) and}} anvar.shukurov@ncl.ac.uk} (AS)}
\author{Anvar Shukurov}
\sameaddress{1}
\begin{abstract}
Recent radio polarization observations have revealed a ple\-tho\-ra of
unexpected features in the polarized Galactic radio background that
arise from propagation effects in the random (turbulent) interstellar
medium. The canals are especially striking among them, a random network
of very dark, narrow regions clearly visible in many directions against
a bright polarized Galactic synchrotron background. There are no obvious
physical structures in the ISM that may have caused the canals, and so
they have been called Faraday ghosts. They evidently carry information
about interstellar turbulence but only now is it becoming clear how this
information can be extracted. Two theories for the origin of the canals
have been proposed; both attribute the canals to Faraday rotation, but
one invokes strong gradients in Faraday rotation in the sky plane
(specifically, in a foreground Faraday screen) and the other only relies
on line-of-sight effects (differential Faraday rotation). In this review
we discuss the physical nature of the canals and how they can be used to
explore statistical properties of interstellar turbulence. This opens
studies of magnetized interstellar turbulence to new methods of
analysis, such as contour statistics and related techniques of
computational geometry and topology. In particular, we can hope to
measure such elusive quantities as the Taylor microscale and the
effective magnetic Reynolds number of interstellar MHD turbulence.
\end{abstract}
\maketitle

\section{The promise and pitfalls of radio polarization}

\begin{figure}[htbp]
\begin{center}
\includegraphics[width=0.98\textwidth]{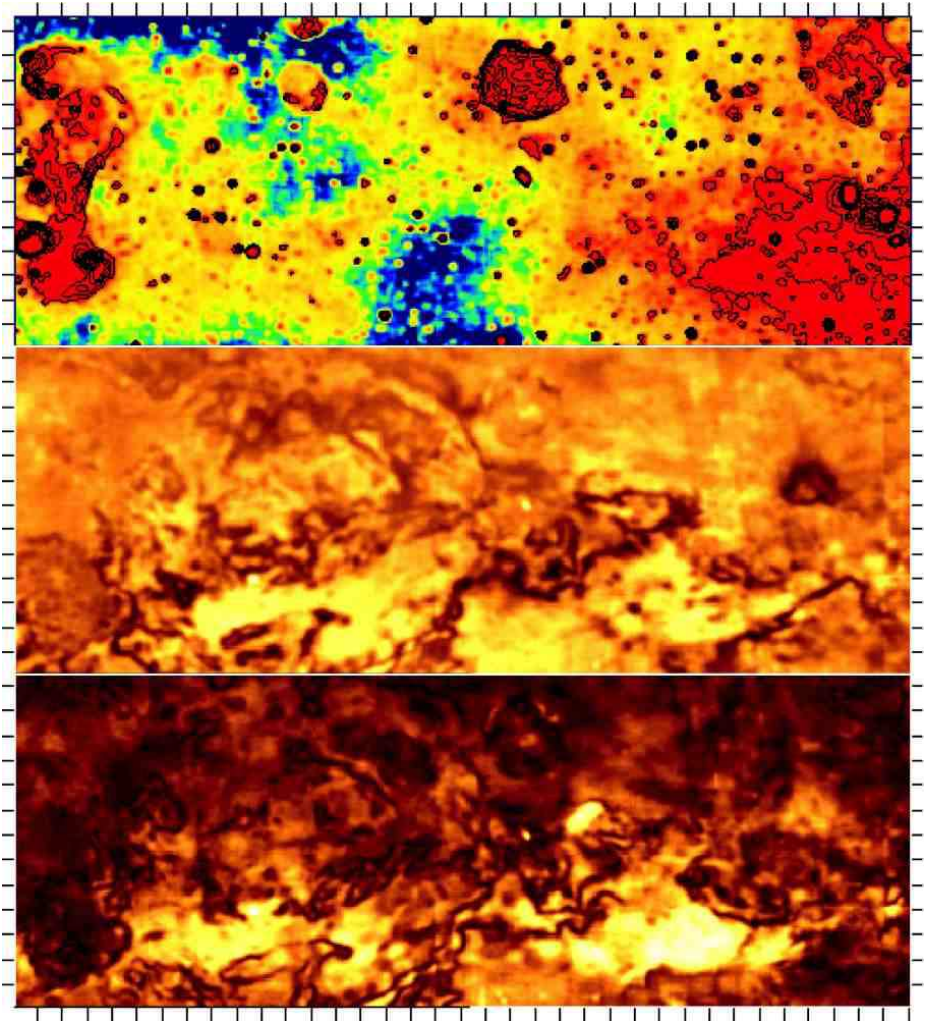}
\caption{A $24\degr\times 9\degr$ section of the $1.4\GHz$ Effelsberg Medium
Latitude Survey centred at $(l,b) = (162\degr, 0\degr)$ (Reich \etal\
\cite{Reich04}). \emph{Top:} total intensity, with large-scale structure added
from Dwingeloo data. \emph{Middle:} polarized intensity, also including the
Dwingeloo large-scale structure. \emph{Bottom:} polarized intensity observed
with the Effelsberg telescope with large scale structure missing. Note the
strong difference between the (preliminary) absolutely calibrated polarized
intensity map (\emph{middle}) and the map with artificial  base-levels
(\emph{bottom}).}
\label{fig:emls}
\end{center}
\end{figure}

High-resolution maps of the Milky Way synchrotron emission show a complex,
irregular system of polarized and depolarized structures (Wieringa \etal\
\cite{Wieringa93}, Uyan{\i}ker \etal\ \cite{Uyaniker98a}, Duncan \etal\
\cite{Duncan99}, Gray \etal\ \cite{Gray99}, Haverkorn \etal\
\cite{Haverkorn00}, Gaensler \etal\ \cite{Gaensler01}, Wolleben \etal\
\cite{Wolleben06}). Many polarized structures can be attributed to
objects in the interstellar medium (ISM), such as supernova remnants and shock
fronts. These objects are usually bright in total intensity as well. However
extended, diffuse polarized emission is also rich in structure which most
plausibly arises from propagation effects as the radiation passes through the random ISM. As illustrated in
Fig.~\ref{fig:emls}, much of this structure is not mirrored in the total
synchrotron intensity so the structure in polarization is not primarily caused
by variations in emissivity. At low frequencies, the cloudy appearance of
polarized intensity maps differs between adjacent, narrow frequency bands
(Haverkorn \etal\ \cite{Haverkorn03}) strongly suggesting that the
polarization varies with wavelength and so a connection with Faraday
rotation is plausible. Random fluctuations in the magnetized ISM
--- caused by its turbulent flows, for example --- leave their
fingerprint in the random appearance of polarized intensity maps. If a connection
can be established between measurable properties of the radio maps to the
physics of the ISM, a powerful new source of information about the dynamic
state of the interstellar plasma will become available, particularly its
little studied magnetic properties. Similarly, a reliable separation of the
Milky Way synchrotron polarized foreground from cosmological signals, either
through the creation of templates from low-frequency data or through
statistical analysis, requires a thorough understanding of the origins of
structure in the diffuse polarized emission. In the context of cosmological
studies, the effects of the ISM on the polarized emission propagating through
it are especially important.

One of the most striking features of the maps is the twisting network of
narrow, dark canals running through regions of bright polarized intensity,
clearly visible in Fig.~\ref{fig:emls} at $1.4\GHz$ and common in other maps
at this frequency (Uyan{\i}ker \etal\ \cite{Uyaniker98a}, Gaensler \etal\
\cite{Gaensler01}) and longer wavelengths (Haverkorn \etal\
\cite{Haverkorn00}).  The observed canals have the following properties:
\begin{enumerate}
\item the observed polarized intensity
falls to zero,
because the polarization angle changes by $90\degr$ across the canal;
\item a canal is one telescope beam wide;
\item a canal passes through a region of significant polarized intensity; 
\item the canal is not related to any obvious structure in the total intensity.
\end{enumerate}

Care in interpreting the observations is especially important in the
case of polarized emission: interferometer observations miss emission at
large scales (determined by their shortest baseline), and single-dish
surveys often have arbitrarily set base-levels around their edges.
Correcting for the missing large-scale structure involves adding a
smoothly varying signal to the Stokes parameters. In the case of total
intensity (Stokes $I$) the consequences are easily predictable. However,
the Stokes parameters $Q$ and $U$ are not positive definite and
therefore algebraic addition of a smooth component can create and/or
remove zeros in either. Hence minima in polarized intensity
$(Q^2+U^2)^{1/2}$ can become maxima and vice versa. The effect of
absolute calibration on the polarization angle
$\Psi=\frac12\arctan{(U/Q)}$ can be equally counter-intuitive. Reich
\etal\ (\cite{Reich04}) clearly identify and discuss these problems. The
dramatic effect the missing data can have on the appearance of the
polarization pattern can be seen by comparing the bottom two panels of
Fig.~\ref{fig:emls}. It is clear that unless statistical parameters can
be identified that are independent of the $Q$ and $U$ intensity values,
polarization surveys should be tied to an absolutely calibrated
reference frame (Uyan{\i}ker \etal\ \cite{Uyaniker98b}, Reich \etal\
\cite{Reich04}). The recent absolutely calibrated all-sky survey at
$1.4\GHz$ by Wolleben \etal\ (\cite{Wolleben06}) will be a mine of
useful information for polarization studies. Haverkorn \etal\
(\cite{Haverkorn04}) argue that if fluctuations in Faraday rotation
measure are sufficiently strong and of a small scale, the resulting
structure in $Q$ and $U$ can be of such a small scale that the missing
short spacings will not cause a problem. In other words, $Q$ and $U$ may
have no large scale structure; this can easily happen at metre
wavelengths.

Two theories accounting for the origin of the canals have been proposed:
one invokes strong plane-of-sky gradients or discontinuities in Faraday
rotation measure in a foreground screen between the emitting region and
the observer (Haverkorn \etal\ \cite{Haverkorn00,Haverkorn04}; Haverkorn
\& Heitsch \cite{Heitsch04}) and the other proposes differential Faraday
rotation along the line-of-sight through the emitting layer (Beck
\cite{Beck99}, Shukurov \& Berkhuijsen \cite{Shukurov03}).

In this review we summarize recent work on the interpretation of
depolarized canals and provide a more detailed explanation of some
analytic results than is available in the readily accessible literature.
After introducing basic equations and defining our notation in
Section~\ref{sec:compP}, we describe the effects of differential Faraday
rotation and Faraday screens, demonstrate how canals are formed by the
two proposed mechanisms and discuss how observable quantities behave in
the vicinity of a canal in each case. Then in Section~\ref{sec:interp}
we turn to the interpretation of the properties of the canals in terms
of the physical state of the ISM. In Section~\ref{subsec:cont} the case
of differential Faraday rotation is discussed in the context of the
statistical properties of contours of a random field and we show how
canals of this type can yield information about the smallest scales of
ISM turbulence. Similar techniques can be applied to canals produced by
Faraday screens, and more generally to contours of any randomly
distributed observable quantity. Section~\ref{subsec:screen}
demonstrates that true discontinuities in Faraday rotation are required
if a foreground screen is to produce canals similar to those observed.
In the case that these discontinuities are shock fronts originating from
supernovae, we show in Section~\ref{subsec:shocks} how the distance
between canals is related to the separation of shock fronts.

\section{The complex polarization, Stokes parameters and depolarization}
\label{sec:compP}

The complex polarization is commonly written as
\begin{equation}
\cP=\po\frac{\int_{V}W(\vect{r})\esyn(\vect{r})\exp{[2i\ps(\vect{r})]\,\dif V}}{
\int_{V}W(\vect{r})\esyn(\vect{r})\ \dif V},
\label{eq:P}
\end{equation}
where $\po\simeq 0.7$ is the maximum degree of polarization for synchrotron
emission (a function of the spectral index), $W$ is the beam profile around a
given position in the sky, $\esyn$ is the synchrotron emissivity, and $\ps$ is
the local polarization angle. The integrals are taken over the
volume of the beam cylinder and $\vect{r}=(x,y,z)$ specifies a location with
respect to the sky plane ($x,y$) and the line of sight, $z$.
The modulus and argument of $\cP$ are the observed degree of polarization $p$
and polarization angle $\Psi$ respectively,
\begin{equation}
\label{eq:comP}
\cP=p\,\exp{(2i\Psi)}.
\end{equation}
When polarized radio emission propagates through magnetized and ionized ISM,
the local polarization angle $\ps$ undergoes Faraday rotation:
\begin{equation}
\ps(\vect{r})= \pso(\vect{r}) + \phi(\vect{r})\;,
\label{eq:psi}
\end{equation}
where $\pso$ is the intrinsic polarization angle (perpendicular to the magnetic
field component transverse to the line of sight at the point of emission),
\begin{equation}
\phi=K\lam^2\int^{\infty}_{z}\nel(z') B_z(z')\, \dif z',
\label{eq:rm}
\end{equation}
$K=0.81 \radmm\p^{-1}\cm^{3}\mkG^{-1}$ is a constant,
$\nel$ is the density of free thermal electrons, $B_z$ is the component of the
magnetic field along the line of sight $z$, $\lam$ is the wavelength and the
observer sits at $z=\infty$. The Faraday depth,
at a position $(x,y)$ in the plane of the sky, is defined as,
\[
F(x,y)=\phi(-\infty)=K\lam^2\int^{\infty}_{-\infty}\nel(z) B_z(z)\, \dif z\;.
\]
and $\phi(z)$ is known as the Faraday depth to a position $z$ (cf.\
Eilek \cite{E89}).
$F$ and $\phi(z)$ are convenient variables to work with; $\phi(z)$ gives
the change in polarization angle of a photon of wavelength $\lam$
passing through the region $z<z^\prime<\infty$, and $F$ is the maximum amount
of Faraday rotation in a given direction in the sky. Note that $F/\lam^2$ can
differ strongly from the observed Faraday rotation measure
$\RM=\dif\ps/\dif\lam^2$. For example, $\RM=F/2\lam^2$ when emission and
rotation occur together in a uniform slab (see below). However, $\RM=F/\lam^2$
for a layer that only rotates --- a Faraday screen.  A more
comprehensive discussion of depolarization and the effects of Faraday rotation
can be found in, e.g., Burn (\cite{Burn66}) and Sokoloff \etal\ (\cite{Sokoloff98}).

Observations of linearly polarized emission measure the Stokes parameters $I$,
$Q$, $U$ and these are related to $\cP$, $p$ and $\Psi$:
\begin{equation}
\cP=\frac{Q+iU}{I}, \quad  p = \frac{\sqrt{Q^2+U^2}}{I}, \quad
\Psi = \sfrac{1}{2}
        \left[\Arctan{\frac{U}{Q}}-\sfrac{1}{2}\pi(\sign Q-1)\sign U\right]\
\label{eq:pqu}
\end{equation}
where $|\Arctan x|<\pi/2$ denotes the principal value; the above form of $\Psi$
allows for a consistent choice of the branch of arctan for any combination
of signs of $Q$ and $U$, so that $-\pi/2\leq\Psi\leq\pi/2$.

\subsection{Differential Faraday rotation}
\label{subsec:diffrot}

\begin{figure}
\begin{center}
\includegraphics[width=0.8\textwidth]{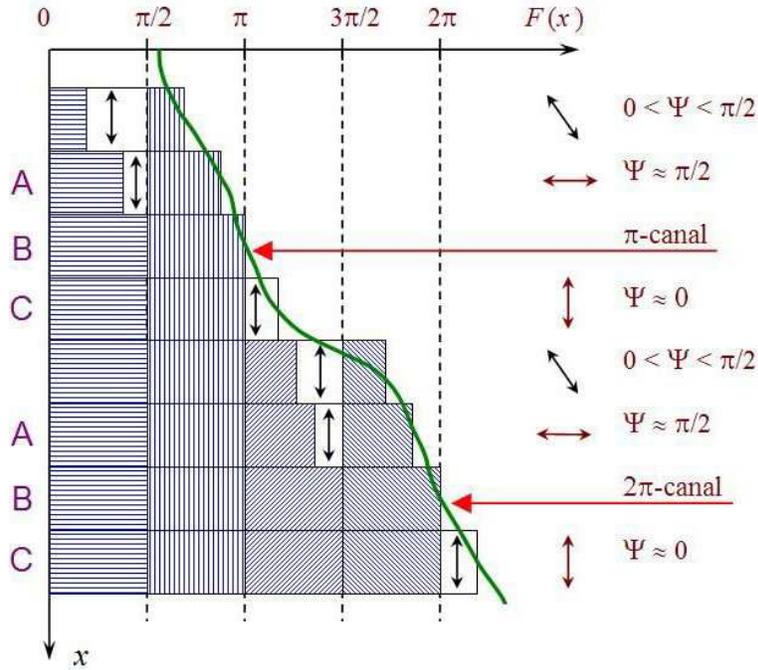}
\caption{Diagram showing how canals are formed by differential Faraday
rotation. The observer is on the right and $x$ is oriented perpendicular to
their line of sight; each horizontal strip represents one telescope beam. The
Faraday depth varies in the plane of the sky (with $x$) as shown with thick solid
line; the intrinsic polarization angle is uniform (vertical double arrows on
the left). Emission generated on the $x$-axis cancels the emission generated
in a layer where the former is rotated by $\pi/2$, i.e., where $F(x)=\pi/2$.
Layers whose emissions mutually cancel are indicated by perpendicularly
hatched boxes. Only the emission from unshaded layers reaches the observer.
The arrows on the right show the observed polarization plane which has been
Faraday rotated in layers to the right of the visible (unshaded) layer to
give the polarization angle $\Psi$ in the range indicated. Note that layers to
the right of the `source' (an unshaded layer) act as a Faraday screen, and so
the rotation angle is equal to $F$. The canals occur where the polarized
emission is fully cancelled (regions labelled B). The emission observed at
values of $F(x)$ slightly smaller than $n\pi$ (regions A) and slightly larger
than $n\pi$ (regions C) originates in different layers. The difference in
$F(x)$ between regions A and C is $\pi/2$. Therefore, emission from region A
is rotated by $\pi/2$ before reaching the observer, whereas emission from
region C is rotated only slightly. Thus, the polarization angles observed on
the two sides of a canal differ by $\pi/2$.  A canal that occurs when
$F(x)=n\pi$ with $n=1$ is shown near the top, and with $n=2$, near the
bottom. (Adapted from Shukurov \& Berkhuijsen \cite{Shukurov03}.)}
\label{fig:diffrot}
\end{center}
\end{figure}

If linearly polarized emission undergoes Faraday rotation, the photons
originating at different positions along a single line of sight will suffer
different rotations of their polarization planes. So even if all the photons
are emitted with the same polarization plane, and the amount of Faraday
rotation per unit path length is constant, the emerging radiation will
comprise a mixture of polarization angles and the resulting degree of
polarization will be reduced. This process is called differential Faraday
rotation (Fig.~\ref{fig:diffrot}); the depolarizing effect is
wavelength-dependent but occurs even for infinitely small telescope beams.

\begin{figure}
\begin{center}
\subfigure[Differential Faraday rotation]{
        \label{fig:geom:diffrot}
        \includegraphics[width=0.35\textwidth]{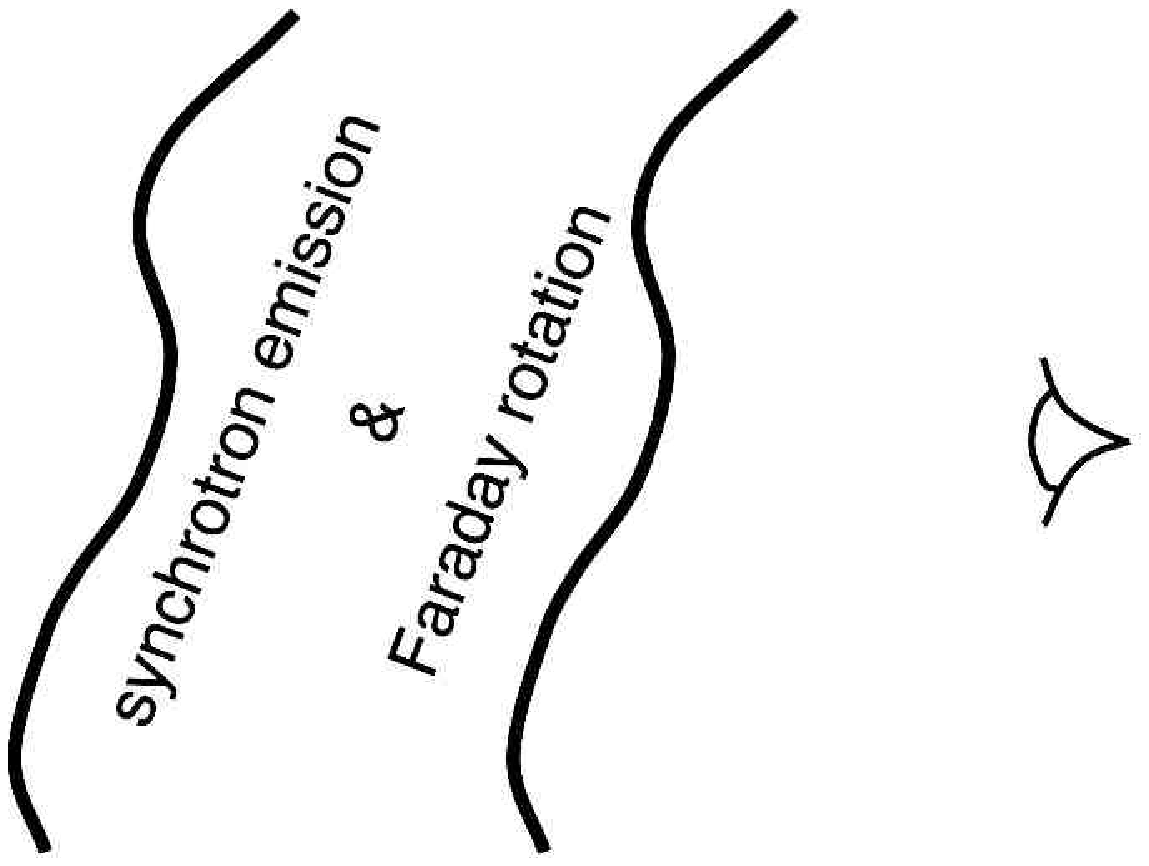}}
\qquad\qquad
\subfigure[Faraday screen]{
        \label{fig:geom:screen}
        \includegraphics[width=0.35\textwidth]{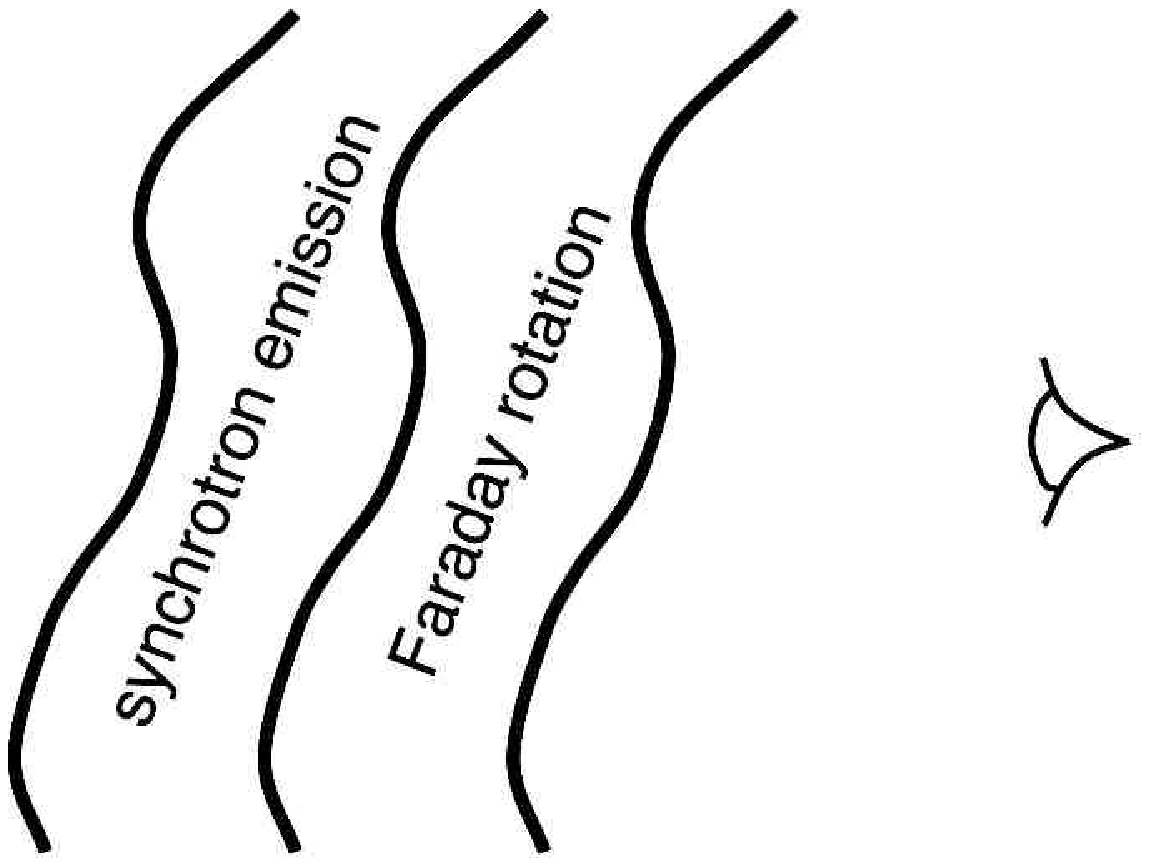}}
\caption{The relative positions, along the line of sight, of the regions
producing synchrotron emission and Faraday rotation. The observer is on the
right. {\bf(a)}: the case of differential Faraday rotation, \ie, both emission
and rotation occur in the same region. {\bf(b)}: a Faraday screen, where all the
emission is produced behind the rotating layer. }
\label{fig:geom}
\end{center}
\end{figure}
\begin{figure}[htbp]
\begin{center}
\subfigure[Differential Faraday rotation]{
	\label{fig:qu:diffrot}
	\includegraphics[width=0.45\textwidth]{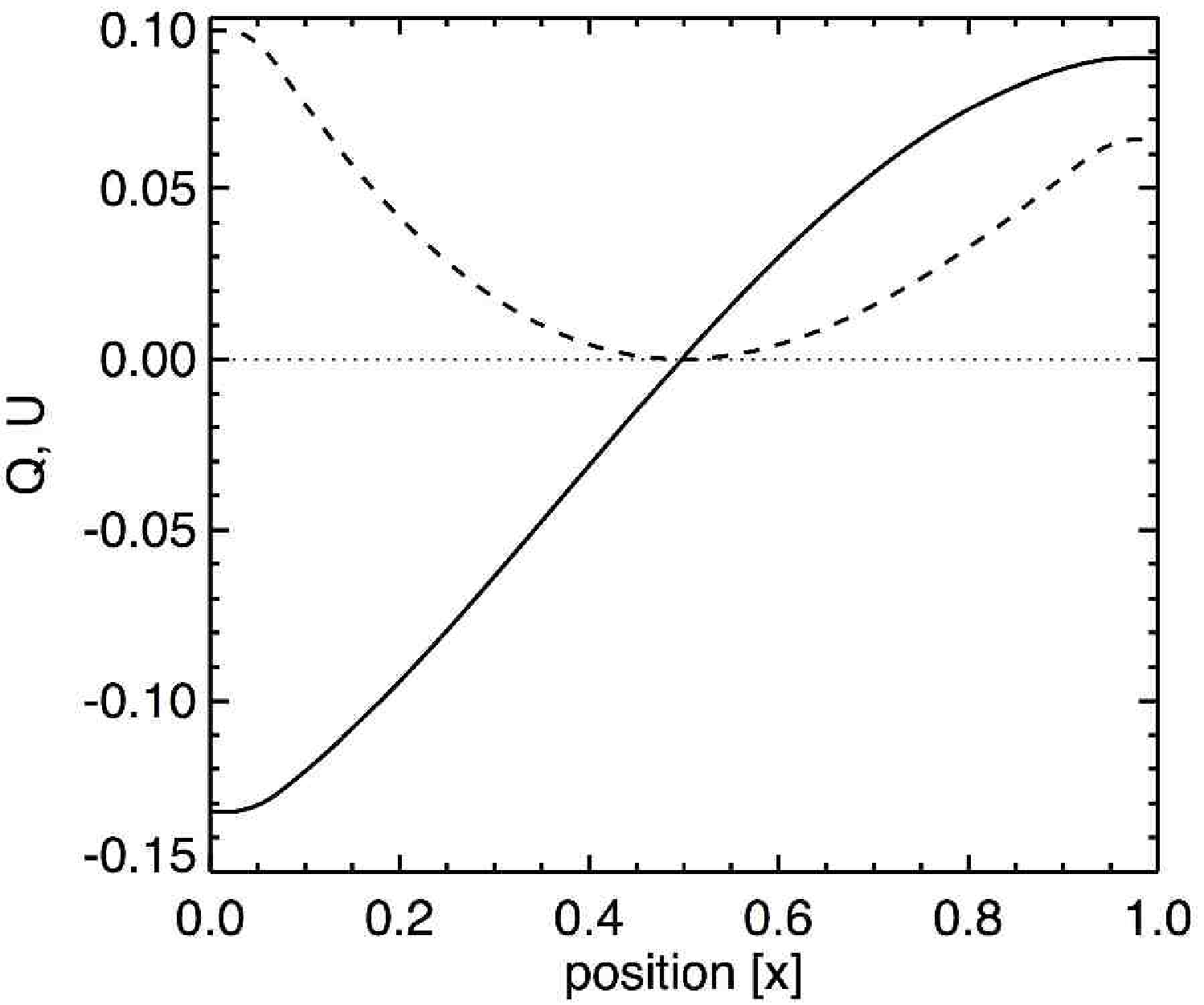}}
\subfigure[Faraday screen]{
	\label{fig:qu:screen}
	\includegraphics[width=0.44\textwidth]{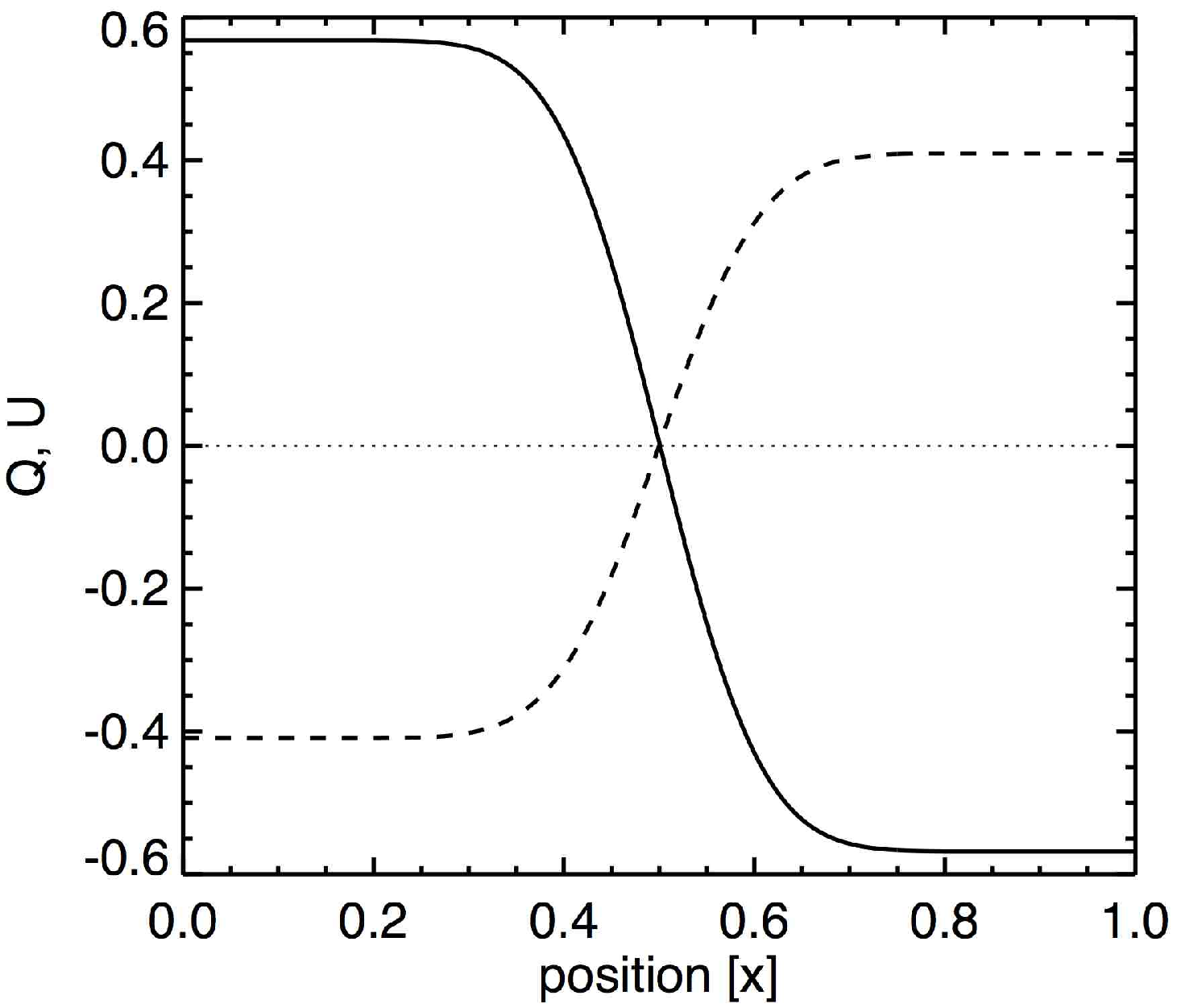}}
\caption{Variation of $Q$ (solid) and $U$ (dashed) (in arbitrary units) across
depolarization canals caused by differential Faraday rotation (a) and
in a nonuniform Faraday screen (b). In each panel, the
canal occurs at $x=X_\mathrm{c}=0.5$ where $Q=U=0$. In Panel (a),
the beam is infinitely narrow and the Faraday
depth varies linearly with $x$ smoothly passing through $F=\pi$ at $x=0.5$.
For the Faraday screen, the beam is a Gaussian of FWHM $D=0.2$,
with the Faraday depth changing discontinuously by $\DF=\sfrac12\pi$
at $x=0.5$. The Stokes parameters $Q$ and $U$ were obtained from Eqs.~(\ref{eq:P})--(\ref{eq:pqu})
for the respective configurations shown in Fig.~\ref{fig:geom}
with $\epsilon,\ \nel B_z$ and $\psi_0$ uniform along the line of sight.
}
\label{fig:qu}
\end{center}
\end{figure}

Although the effect appears rather simple to understand, one can
discover unexpected subtleties from simple mathematical analysis. Consider a
slab extending between $-h<z<h$ along the line of sight, with constant
synchrotron emissivity, $\epsilon=\mbox{const}$, and uniform magneto-ionic medium,
$\nel B_z=\mbox{const}$, throughout
as shown in Fig.~\ref{fig:geom:diffrot}. The
telescope beam is assumed to be infinitely narrow for simplicity, i.e. $W$ is
a $\delta$-function in $x,y$; this makes trivial integration in the sky plane
in Eq.~(\ref{eq:P}). Equation~(\ref{eq:rm}) then yields
\begin{equation}
\label{eq:F}
\phi(z)=\left\{
\begin{array}{ll}
F\;,                  &z<-h\;,\\
\sfrac12 F(1-z/h)\;,  &|z|\leq h\;,\\
0\;,                    &z>h\;,
\end{array}
\right.
\end{equation}
where $F=2K\lam^2\nel B_z$ is the Faraday depth of the slab.
Then, for $\pso=\mathrm{const}$,
Eqs.~(\ref{eq:P}) and (\ref{eq:psi}) lead to the well known result
(Burn~\cite{Burn66})
\begin{equation}
\label{eq:diffrot}
\cP=\po\,\frac{\sin{F}}{F}\,\exp{\left[2i\left(\pso+\sfrac{1}{2}F\right)\right]}.
\end{equation}
To see more clearly how differential Faraday rotation produces depolarization
canals, it is useful to rewrite this with the help of Eqs.~(\ref{eq:comP})
and (\ref{eq:pqu}):
\begin{eqnarray}
p & = & \po\left| \frac{\sin{F}}{F}\right|, \label{eq:diffrot:p}\\
\frac{Q}{I} & = & \po\frac{\sin{F}}{F}\cos{2(\pso+\sfrac{1}{2}F)}, \label{eq:diffrot:q}\\
\frac{U}{I} & = & \po\frac{\sin{F}}{F}\sin{2(\pso+\sfrac{1}{2}F)}. \label{eq:diffrot:u}
\end{eqnarray}
Equation~(\ref{eq:diffrot:p}) shows that $p=0$ where $|F|=n\pi$, $n=1,2
\dots$. If $F$ is a continuous function of position with well-behaved
contours, canals will develop on the contours of $F$ where
$|F|=n\pi$. Therefore, for a canal produced by differential rotation we know
the magnitude of the Faraday depth along the canal (up to an integer factor
$n$). Techniques for retrieving further information about the physical state
of the ISM from the statistical properties of these contours is discussed in
Section~\ref{subsec:cont}.

We are free to measure the polarization angle from a line perpendicular to
the transvese magnetic field near the canal; then $\pso=0$. Equations~(\ref{eq:diffrot:q}) and
(\ref{eq:diffrot:u}) show that in this reference frame
\[
Q\propto\sin{F}\,\cos{F}\;,\qquad  U\propto\sin^2{F}\;.
\]
When $F(\vect{x})$ passes through an integer multiple of $\pi$, both $Q$
and $U$ vanish to produce a canal, but $Q$ will change sign across the canal and $U$ will
not. In other words, there is always a reference frame in the sky plane
such that one of the Stokes parameters changes sign, but the other does
not across a canal produced by differential Faraday rotation. This
behaviour is illustrated in Fig.~\ref{fig:qu:diffrot} and we will see in
Section~\ref{subsec:screen} that this does not occur in the case of
canals formed in a Faraday screen as illustrated in
Fig.~\ref{fig:qu:screen}.

For the signs of $Q$ and $U$ shown in Fig.~\ref{fig:qu:diffrot}, $\Psi$ is in
the second quadrant to the left of the canal and in the first quadrant on the
right of it. As $Q\to0$ and $U\to0$ for $x\to X_\mathrm{c}$, where
$X_\mathrm{c}$ is the position of the canal, we have $\Psi=-\pi/2$ for
$x=X_\mathrm{c}-0$ but $\Psi=\pi/2$ for $x=X_\mathrm{c}+0$ --- see also
Eq.~(\ref{eq:pqu}). Thus the polarization angle changes discontinuously across
the canal although the magneto-ionic medium is perfectly continuous. In particular,
$\RM$ has the same values on both sides of a canal (but is undefined at it).
As illustrated in Fig.~\ref{fig:diffrot}, the jump of the polarization angle
in a continuous medium occurs because the polarized photons detected on the
either side of the canal originate at different depths in the source and
therefore have experienced different amounts of Faraday rotation.

The discussion above refers to the case of an infinitely narrow telescope
beam; these canals are infinitely narrow curves in the sky plane. The effect
of a finite beam width is twofold. Firstly, the canals in observed maps will
be one beam wide, as the telescope combines emission from either side of the
contour where $|F|=n\pi$. Secondly, the canals will gradually fill up with
polarized emission as the resolution coarsens, so that asymmetry of the
magneto-ionic medium within the beam becomes noticeable.

Canals are also filled up if the synchrotron emissivity varies along the line
of sight provided this variation is not compensated by a variation in Faraday
rotation. It can be shown, in particular, that a double-layered source where
synchrotron intensities produced in each layer are $I_1$ and $I_2$, with
$|I_1-I_2|/I_1=\varepsilon$, and $F=n\pi$, will produce a partially filled
canal with the fractional polarization $p/p_0\simeq\varepsilon/\pi$, valid for
$\varepsilon\ll1$, \ie, for weak inhomogeneity. This effect can
explain why the observed canals are not all closed curves as might be expected
of contours of a continuous function.

Tangled magnetic fields and fluctuations of the electron number density can
produce additional depolarization known as internal Faraday dispersion,
characterized by the standard deviation of the Faraday depth, $\sigma_F$. This
effect also fills up a canal, so that the minimum fractional polarization is
$p/p_0\simeq \sigma_F^2/\pi$ for $\sigma_F^2\ll\pi$. The above two effects are
discussed by Fletcher \& Shukurov (in preparation).

Joint action of various depolarization effects can produce counter-intuitive
effects. For example, Sokoloff \etal\ (\cite{Sokoloff98}, their Sect. 6.2 and
Fig.~5 in Erratum) show that internal Faraday dispersion combined with
differential Faraday rotation can lead to a jump of $90\degr$ in polarization
angle whilst the degree of polarization varies smoothly and $p$ differs from
zero (if observed with infinitely narrow beam); this would appear as
a canal when observed at finite resolution because of the jump in the
polarization angle.

The position of a canal produced by differential Faraday rotation must change
with the wavelength at which it is observed because $F$ depends on $\lambda$
according to Eq.~(\ref{eq:rm}). To a first approximation, a change of wavelength
from $\lambda$ to $\lambda'=\lambda+\delta\lambda$ will displace the canal by $\delta x$ where
\[
\delta x \simeq \left|\frac{\delta F}{|\nabla F|}\right|\;,
\qquad \delta F = F(\lambda')-F(\lambda),
\qquad |\delta\lambda|\ll\lambda\;.
\]
For example, if differential Faraday rotation is the only depolarization
mechanism, we have $F=2\RM\,\lambda^2$ with $\RM$ being independent of
$\lambda$, and then
\[
\delta x \simeq 2L_\RM \left|\frac{\delta\lambda}{\lambda}\right|\;,
\qquad
L_\RM=\left|\frac{\RM}{|\nabla\RM|}\right|\;,
\]
where $L_\RM$ is the scale at which $\RM$ varies.
Shukurov \& Berkhuijsen (\cite{Shukurov03}) argue, using a similar estimate,
that the displacement can be difficult to observe if $|\delta\lambda|\la\lambda$,
which is often the case.

\subsection{Foreground Faraday screens}
\label{subsec:screen}

\begin{figure}
\begin{center}
\includegraphics[width=0.5\textwidth]{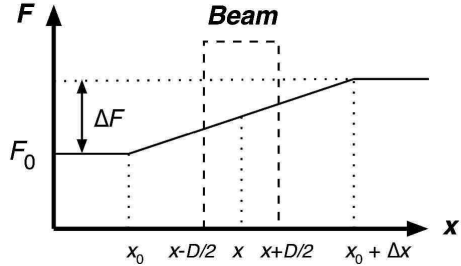}
\caption{Sketch showing the variation of Faraday depth $F$ across the sky,
here along $x$. There is a constant gradient of $F$ in the range
$x_0<x<x_0+\Delta x$, and the telescope beam has a width $D$.}
\label{fig:grad}
\end{center}
\end{figure}

If synchrotron emission and Faraday rotation occur in distinct layers --- with
the rotating layer nearer to the observer as illustrated in
Fig.~\ref{fig:geom:screen} --- the foreground magneto-ionic medium is called a
Faraday screen. If Faraday rotation in the screen is uniform across a
telescope beam then no depolarization occurs; the polarization angle of
radiation emerging from the emitting layer is rotated by $\Delta\Psi=F$, but
the mixing of different angles within a beam, that causes depolarization, does
not occur. A Faraday screen can only cause depolarization if it is nonuniform
on the scale of the beam, so that different lines of sight within the
beam undergo differing amounts of Faraday rotation.

\begin{figure}
\begin{minipage}[c]{0.48\textwidth}
	\centering\includegraphics[width=\textwidth]{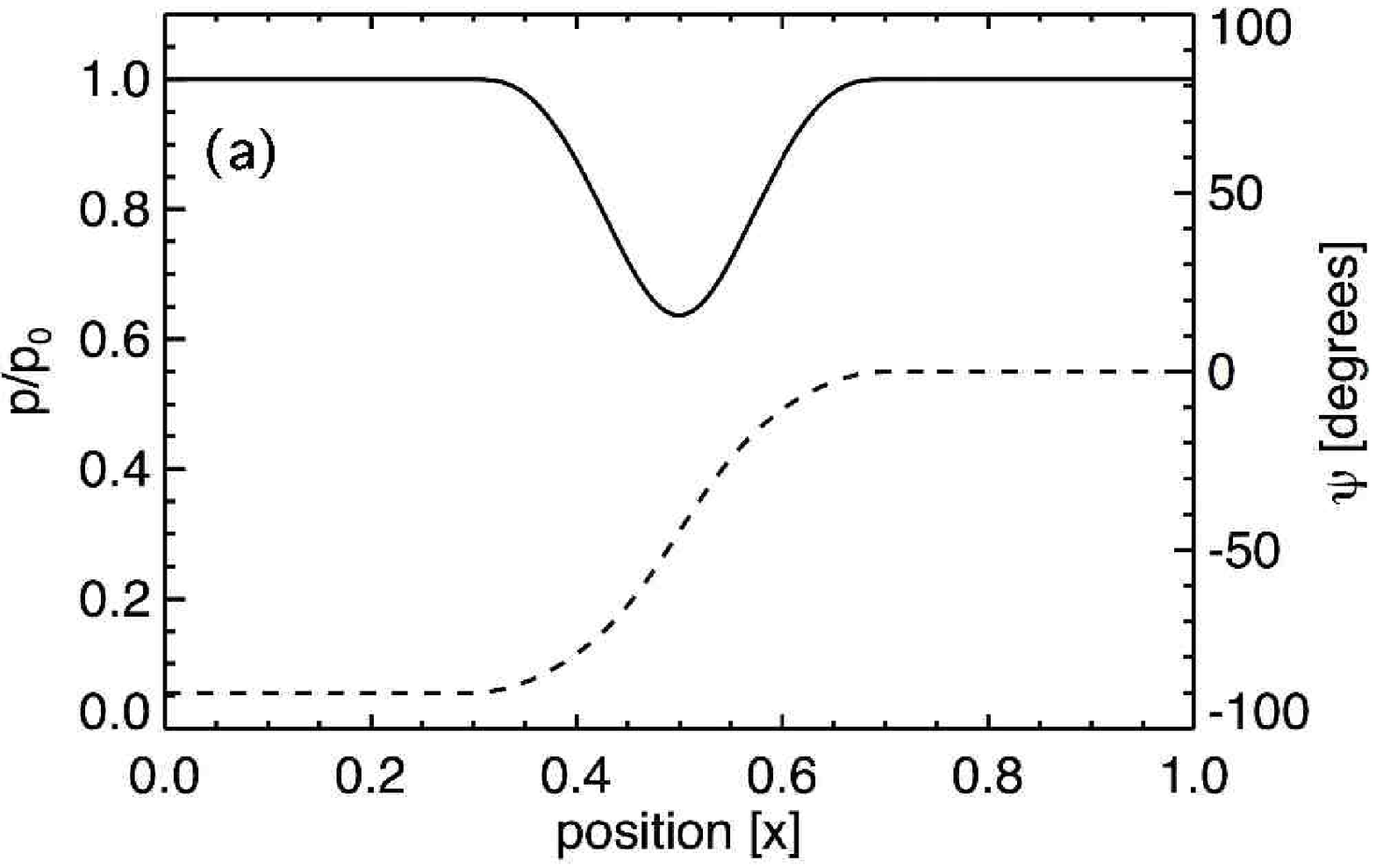}
	\end{minipage}
\begin{minipage}[c]{0.48\textwidth}
	\centering\includegraphics[width=\textwidth]{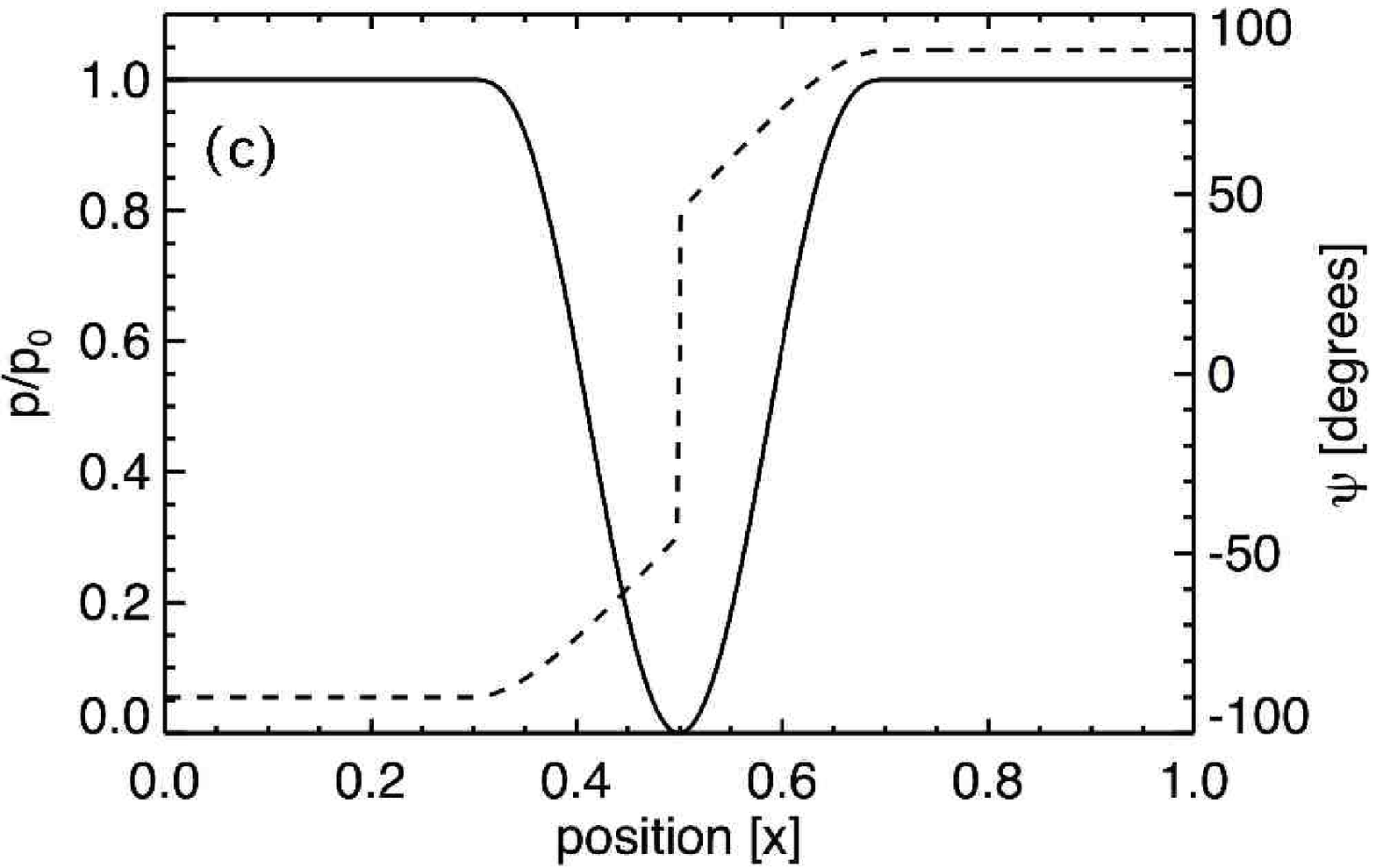}
        \end{minipage}
\vspace{5mm}5
\begin{minipage}[c]{0.48\textwidth}
	\centering
	\includegraphics[width=0.55\textwidth]{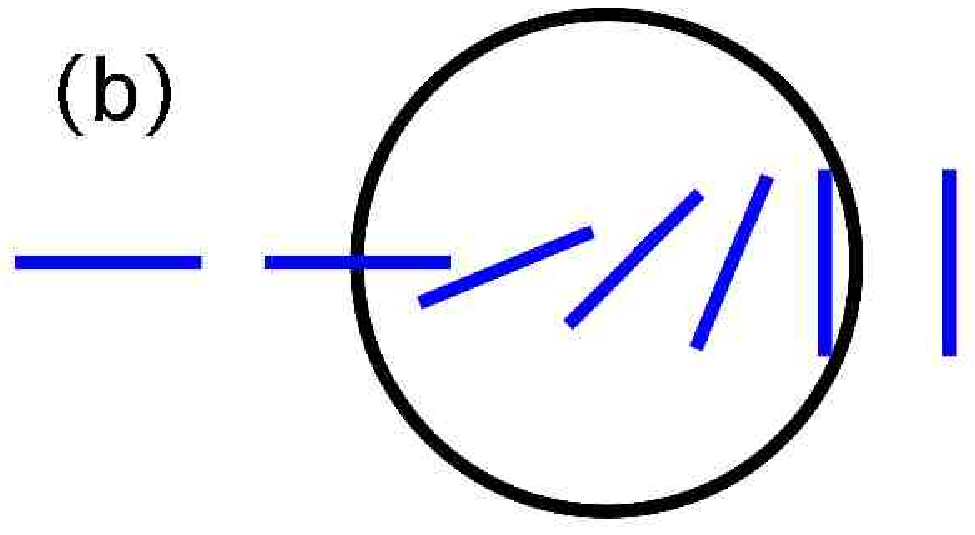}
	\end{minipage}
\begin{minipage}[c]{0.48\textwidth}
	\centering\includegraphics[width=0.55\textwidth]{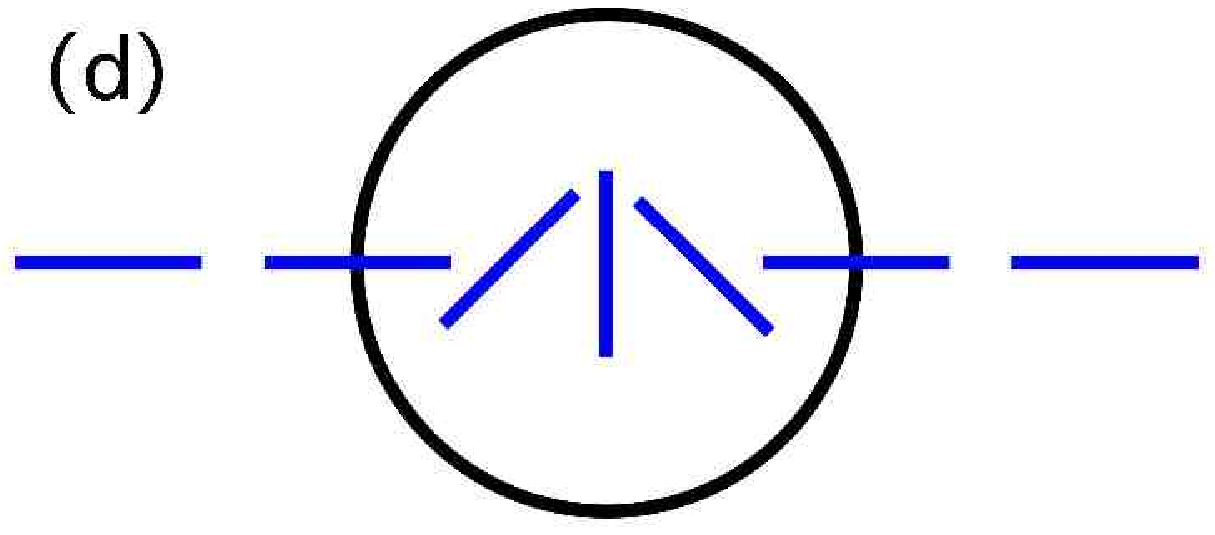}
	\end{minipage}\\
\caption{The variation of the fractional polarization and polarization angle
across a Faraday screen with a smooth variation of Faraday depth described in
Fig.~\ref{fig:grad} with $x_0=0.4$, $\Delta x=0.2$ and $X_\mathrm{c}=0.5$.
Left-hand panels: {\bf(a)} the change in the beam-averaged degree of
polarization (solid) and polarization angle (dashed) for $\Delta F=\pi/2$
convolved with a square beam of width $D=0.2$. The fractional
polarization does not reduce to zero although $\Psi$ changes by $\pi/2$ across
the region. The reason is illustrated in panel {\bf(b)} where we show the
change in the polarization angles obtained along individual lines of sight
(dashes) within the beam (circle). It is clear that the polarization cannot
drop to zero in this configuration. The right-hand panels show the same as on
the left, except that here $\DF=\pi$. Now, panel {\bf(c)} shows that the
fractional polarization does drop to zero at $x=X_\mathrm{c}$ producing a
canal, but the corresponding change in $\Psi$ across the canal is $\pi$. Panel
 {\bf(d)} explains how this variation in $\Psi$ among individual lines of
sight leads to the complete depolarization. The calculation used
Eqs.~\ref{eq:P}--\ref{eq:pqu} for the configuration of Fig.~\ref{fig:grad}.}
\label{fig:screen}
\end{figure}

Consider a uniform polarized radio source shining through a
Faraday rotating layer. Consider a simple variation of the Faraday depth in
the sky plane shown in Fig.~\ref{fig:grad}, where $F$ changes by
$\Delta F$ across a distance $\Delta x$, and the width of the telescope beam is
$D$. We assume that the telescope beam has a square profile for simplicity. In the regions where $F=\mathrm{const}$ in
Fig.~\ref{fig:grad}, Eq.~(\ref{eq:P}) reduces to
\begin{equation}
\label{eq:screen:sides}
\cP=\po\left\{
\begin{array}{ll}
\exp[2i(\pso+F_0)]\;,        &x<x_0-D/2\;, \\
\exp[2i(\pso+F_0+\DF)]\;,    &x>x_0+\Delta x+D/2\;,
\end{array}
\right.
\end{equation}
so $p=\mathrm{Re}\,\cP=\po$ and no depolarization occurs if $F$ is constant.
In the region
$x_0+D/2<x<x_0-D/2+\Delta x$, where $F$ varies linearly, Eq.~(\ref{eq:P}) yields
\begin{eqnarray}
\label{eq:screen}
\cP & = & \po\exp[2i(\pso + F_0)]\frac{1}{D} \int^{x+D/2}_{x-D/2}
        \exp{\left(2i \DFD\frac{x-x_0}{D}\right)}\,\dif x \nonumber \\
 & = & \po\,\frac{\sin{\DFD}}{\DFD}\, \exp{[2i(\pso+F)]}\;,
\end{eqnarray}
where $\DFD=D\,\dif F/\dif x$ is the increment in the Faraday depth across the
telescope beam arising from a continuous gradient in $F$.  (We ignore, for simplicity,
the ranges $|x-x_0|<D/2$ and $|x-x_0+\Delta x|<D/2$ with a mixture of constant $F$
and a gradient in $F$ within a beam.)
Then
\begin{equation}
\label{eq:screen:p}
p = \po\left| \frac{\sin{\DFD}}{\DFD}\right|.
\end{equation}
Thus, the depolarization in a continuously inhomogeneous screen is
complete if $|\DFD|=n\pi$ and we can then expect a canal to form along a
contour where $|\nabla F|=n\pi/D$ (where we have replaced $\dif F/\dif
x$ by $\nabla F$). An increment of $|\DFD|=n\pi$ across the beam means
that the change in polarization angle across the canal is
$\Delta\Psi=180\degr,$ \ie, no change --- see Eq.~\ref{eq:screen}.
However, canals attributed to a Faraday screen by Haverkorn \etal\
(\cite{Haverkorn00}) have $\Delta\Psi=90\degr$. As we illustrate in
Fig.~\ref{fig:screen}, this change in $\Psi$ in fact cannot produce a
canal if $F$ varies continuously: a gradient in $F$ that produces a
$90\degr$ change in polarization angle across the beam does not lead to
complete depolarization, whereas a gradient that produces $180\degr$ of
Faraday rotation does give $p=0$. Canals with $\Delta\Psi=0$ (or
$180\degr$) predicted here have not yet been observed.

Depolarization can be complete if the change in Faraday rotation within the
beam produces a mixture of polarization angles that exactly cancel each other.
Complete cancellation depends on both the variations in the Faraday depth
across the screen $\DF$ and the telescope resolution, so that the
pattern of canals can change very significantly if observed at different resolution.

\begin{figure}
\begin{center}
\includegraphics[width=0.6\textwidth]{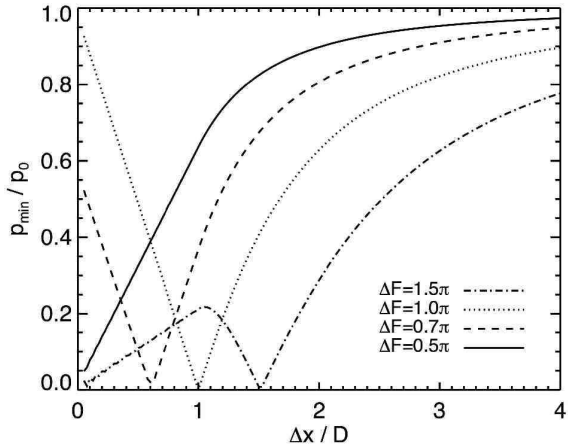}
\caption{The minimum degree of polarization produced by a gradient of the
Faraday depth in a foreground screen, $p_\mathrm{min}$, as a function of the
relative extent of the gradient $\Delta x/D$, where $D$ is the width of a
square beam. Different curves represent different Faraday rotation increments
($\DF$, shown in legend). For $\Delta x/D<1$, $F$ varies over
a range narrower than the beam width, and is effectively discontinuous when
$\Delta x/D\ll1$. Complete depolarization in a discontinuous screen, $\Delta x/D\to0$, occurs if
$\DF=(2n+1)\pi/2,\ n=0,1,\ldots$.}
\label{fig:pmin}
\end{center}
\end{figure}

Figure~\ref{fig:pmin} shows how the depth of a canal produced by a
Faraday screen depends on the increment in Faraday depth across the beam
$\DFD$ and on how well resolved this increment is, $\Delta x/D$. We can
see that \emph{all} gradients in $F$ can produce $p=0$ if observed at an
appropriate resolution and so we should expect to see canals with
\emph{all\/} jumps in polarization angle across them.  The only
condition for that is that the total variation in $F$ available exceeds
$\pi/2$, the minimum that can produce a canal for $\Delta x/D\ll1$. The
jump in $\Psi$ across the canal (which is one beam wide) in Fig.~\ref{fig:pmin} is equal to the
corresponding value of $\DF$ (shown in the legend) if $\Delta x/D\leq1$
and to $\DFD$ otherwise.

It is evident from Fig.~\ref{fig:pmin} that the only way that canals with
$\Delta\Psi=90\degr$ will be preferentially created is if the variations
in $F$ in the Faraday screen are \emph{discontinuous\/}. In this case
complete cancellation will only occur if all of the polarization angles
within a beam are a mixture of exactly orthogonal pairs and  canals will
be formed along contours where $\DFD=(2n+1)\pi/2$. Like the case of
differential Faraday rotation, canals produced by discontinuities in a
Faraday screen do not depend on the resolution of the observations. In
Section~\ref{subsec:shocks} we will discuss how the statistical
properties of canals formed by discontinuities in a Faraday screen can
provide information about the shock-wave turbulence in the diffuse ISM.

For a canal produced by a discontinuous change of $\DF$, the behaviour of
$Q$ and $U$ across the canal can obtained from Eq.~(\ref{eq:screen:sides}) with
$\DF=(2n+1)\pi/2$: each changes
by $\pi$ (neglecting $2n\pi$ in the arguments of $\cos$ and $\sin$),
\begin{eqnarray}
\label{eq:screen:qu}
\frac{Q}{I}=\po\left\{
\begin{array}{ll}
\cos{[2(\pso+F_0)]}, &x<X_\mathrm{c}\;,\\
\cos{[2(\pso+F_0)+\pi]},        &x>X_\mathrm{c}\;
\end{array}
\right.
\end{eqnarray}
and likewise for $U$.
Figure~\ref{fig:qu} shows an example
of how $Q$ and $U$ change across a canal produced in a Faraday screen and by
differential Faraday rotation. In contrast to the case of differential
Faraday rotation,  Eqs.~(\ref{eq:diffrot:q}) and (\ref{eq:diffrot:u}), \emph{both\/}
$Q$ and $U$ of Eq.~\ref{eq:screen:qu} will \emph{always\/} change sign across a canal produced by a
Faraday screen. (A special case is where $Q$ (or $U$) is zero whereas $U$ (or $Q$) changes sign at the canal.) This offers an opportunity to reveal the nature of an observed canal from
a careful analysis of the variation of $Q$ and $U$ across it.

\section{What can we learn about the ISM from depolarization canals?}
\label{sec:interp}
If the mechanism which forms a canal can be identified --- for example by
looking at the variation of $Q$ and $U$ across the canal --- we then know the
location of level lines of Faraday depth where $|F|=n\pi$, in the case of
differential rotation, or the projected position of discontinuities with
$|\DFD|=(2n+1)\pi/2$, for the action of a Faraday screen. In both cases, canals
represent contours of a random function of position in the plane of the sky,
namely the Faraday depth or the magnitude of its gradient. In this section
we discuss how statistical properties of the contours of a random function can
be used to measure its parameters, and hence how statistical properties of the
depolarization canals can be interpreted in terms of the properties of the
random (turbulent) ISM.

\subsection{Differential Faraday rotation: contour statistics and turbulence}
\label{subsec:cont}
As we show here, the mean separation of the contours of a differentiable
random function is sensitive to the curvature of its autocorrelation
function at zero lag, i.e., to a parameter known as the Taylor
microscale in turbulence theory. Reviews of this theory can be found in
Longuet-Higgins~(\cite{Longuet57}), Sveshnikov~(\cite{Sveshnikov66}) and
Vanmarcke~(\cite{Vanmarcke83}). Observations of canals (or contours of
another observable), obtained under quite a modest resolution, provide a
possibility to measure the Taylor microscale, related to the dissipation
scale of the turbulent flow and to the Reynolds number. Contour
statistics have been used in astrophysics in the studies of structure
formation (Peebles \cite{P84}; Barden \etal\ \cite{B86}; Ryden
\cite{R88}; Ryden \etal\ \cite{R89}) and the cosmic microwave background
(Coles \& Barrow \cite{CB87}).

\begin{figure}
\begin{center}
\includegraphics[width=0.6\textwidth]{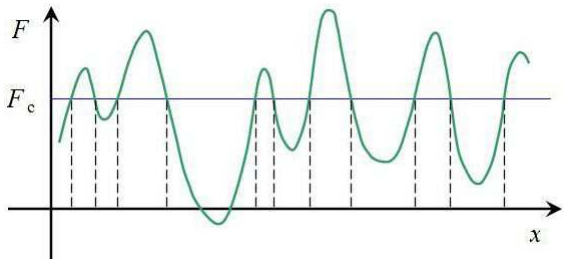}
\caption{Crossings of a level $F_\mathrm{c}$ by a random function $F(x)$ are
indicated with vertical dashed lines.}
\label{crossings}
\end{center}
\end{figure}

\subsubsection{Overshoots of a random function in one dimension}
We start with crossings of a random function of a single variable over a
reference level (Fig.~\ref{crossings}); a clear and detailed discussion
of this problem can be found in \S9 of Sveshnikov (\cite{Sveshnikov66}).
Consider a stationary, differentiable random function of position
$\cF(x)$, whose realisation, or measured value, is denoted by $F(x)$.
(We have in mind applications where $F$ can be the Faraday depth or the
magnitude of its gradient.) The function will pass, from below, through
the level $F_\mathrm{c}$ in the infinitesimal interval $(x,x+\dif x)$ if
the following inequalities are satisfied simultaneously:
$\cF(x)<F_\mathrm{c}$ and $\cF(x+\dif x)>F_\mathrm{c}$. For a continuous
$\cF(x)$, we can write $\cF(x+\dif x)\approx\cF(x)+\cF'(x)\,\dif x$,
where $\cF'=\dif \cF/\dif x$. Thus we have
\begin{equation}
\label{range}
F_\mathrm{c}-\cF'(x)\,\dif x < \cF(x) < F_\mathrm{c}\quad \mbox{for }\cF'(x)>0\;.
\end{equation}
In order to calculate the probability that these two inequalities hold,
consider the joint probability density of $\cF$ and its derivative
$\cF'$ at a given position $x$, which we denote $f(F,F')$. Then the
probability of a overshoot above a level $F_\mathrm{c}$ is given by
\[
P\{F_\mathrm{c}-\cF'(x)\,\dif x < \cF(x) < F_\mathrm{c}\} =
\int_0^\infty \dif F' \int_{F_\mathrm{c}-F'(x)\,\dif x}^{F_\mathrm{c}} \dif F\,f(F,F')\;,
\]
where integrations are performed over all values of $\cF$ and $\cF'$ that satisfy
Eq.~(\ref{range}). The integral over $F$ is immediately taken using the mean value
theorem:
\[
\int_{F_\mathrm{c}-F'(x)\,\dif F}^{F_\mathrm{c}} f(F,F')\,\dif x
= f(F_\mathrm{c},F')F'\,\dif x\;,
\]
and then
\begin{equation}\label{prob_over}
P\{F_\mathrm{c}-\cF'(x)\,\dif x < \cF(x) < F_\mathrm{c}\} =\dif x
\int_0^\infty f(F_\mathrm{c},F')F'\,\dif F'\;.
\end{equation}
It is convenient to introduce the probability density of an overshoot,
\ie, the probability $p_\mathrm{a}(F_\mathrm{c})$ of an overshoot above
a level $F_\mathrm{c}$ at a position $x$ per unit interval of $x$. The
above result can be written as
\[
p_\mathrm{a}(F_\mathrm{c})=\int_0^\infty f(F_\mathrm{c},F')\,F'\,\dif F'\;.
\]
Similarly, we obtain the probability density for $F(x)$ to cross the level
$F_\mathrm{c}$ from above:
\[
p_\mathrm{b}(F_\mathrm{c})=-\int_{-\infty}^0 f(F_\mathrm{c},F')\,F'\,\dif F'\;,
\]
and the total probability density for $F(x)$ to cross the level $F_\mathrm{c}$ at a
position $x$ either from above or from below is given by
\begin{equation}\label{pc}
p_\mathrm{c}=p_\mathrm{a}+p_\mathrm{b}
=\int_{-\infty}^\infty f(F_\mathrm{c},F')\,|F'|\,\dif F'\;.
\end{equation}

It can be shown (Sveshnikov \cite{Sveshnikov66}) that the mean numbers
of crossings of a level $F_\mathrm{c}$ per unit length, $N_1$, is equal
to $p_\mathrm{c}$, \ie, the mean separation of the crossing points is
given by
\[
\mean{x}_\mathrm{c}=N_1^{-1}=p_\mathrm{c}^{-1}\;,
\]
where overbar denotes averaging.

The results can be written in a closed, explicit form if $\cF(x)$ is a Gaussian
random function because then the fact that $\cF(x)$ and $\cF'(x)$ are
uncorrelated (which is true for any stationary, differentiable random function)
implies that they are statistically independent, so that
\[
f(F,F')=f_F(F)f_{F'}(F')\;,
\]
and hence (we note that $\mean{F'}=0$)
\[
f(F,F')= \frac{1}{\sigma_F(2\pi)^{1/2}}\,
                \exp{\left[-\frac{(F-\mean{F})^2}{2\sigma_F^2}\right]}\,
         \frac{1}{\sigma_{F'}(2\pi)^{1/2}}\,
                \exp{\left[-\frac{(F')^2}{2\sigma_{F'}^2}\right]}\;,
\]
where we have introduced the standard deviations of $F$ and $F'$, $\sigma_F$ and
$\sigma_{F'}$, respectively. Integration over $F'$ in Eq.~(\ref{pc}) can now
be easily performed to obtain
\begin{equation}\label{probc}
N_1=p_\mathrm{c}=\frac{1}{\pi}\,\frac{\sigma_F}{\sigma_{F'}}
\exp{\left[-\frac{(F_\mathrm{c}-\overline{F})^2}{2\sigma_F^2}\right]}.
\end{equation}
In application to canals, where $F_\mathrm{c}=n\pi,\ n=1,2,\ldots$ in
the case of differential Faraday rotation, one needs to calculate the
mean separation of positions where $|F(x)|=F_\mathrm{c}$. The
corresponding probability density involves an additional contribution
arising from crossings at $F(x)=-F_\mathrm{c}$, which should be added to
the right-hand side of Eq.~(\ref{probc}). The additional term is
proportional to $\exp[-(F_\mathrm{c}+\mean{F})^2/2\sigma_F^2]$. This
term can be significant if $\mean{F}\approx0$, so that
$|F_\mathrm{c}-\mean{F}|\simeq|F_\mathrm{c}+\mean{F}|$. If only one of
the two terms is significant, it is sufficient to replace $\mean{F}$ by
$|\mean{F}|$ in  Eq.~(\ref{probc}).

Since
\[
\frac{\sigma_{F'}^2}{\sigma_F^2}
=-\left.\frac{\dif^2 C(r)}{\dif r^2}\right|_{r=0} =\frac{2}{l_\mathrm{T}^2}\;,
\]
where $C(r)$ is the autocorrelation function of $F(x)$ normalised to $C(0)=1$ and
$l_\mathrm{T}$ is known as the Taylor microscale (e.g., Sect.~6.4 in
Tennekes \& Lumley \cite{TL72}), we obtain the following expression for the
mean separation of the positions where $|F(x)|=F_\mathrm{c}$:
\begin{equation}                      \label{meanx0}
\mean{x}_\mathrm{c}=\frac{\pi}{\sqrt2}\, l_\mathrm{T}\,
\exp{\left[\frac{(F_\mathrm{c}-|\overline{F}|)^2}{2\sigma_F^2}\right]}\;.
\end{equation}

\subsubsection{Contour statistics in two dimensions}
The theory of overshoots can be generalized to two
dimensions, where we consider a random field $\cF(\vect{x})$ depending on two
variables $\vect{x}=(x,y)$ (e.g., position in the sky plane); we also
introduce $\vect{V}=\nabla\cF$. The probability of an overshoot is now equal
to the probability that the following inequalities are satisfied
simultaneously:
\[
\cF(\vect{x})<F_\mathrm{c}\;,\qquad
\cF(\vect{x}')>F_\mathrm{c}\;,
\qquad
\vect{x}'=\vect{x}+\dif\vect{x}\;,
\]
where $\dif\vect{x}=\widehat{\vect{V}}\,\dif x$, with
$\widehat{\vect{V}}=\nabla\cF/|\nabla\cF|$ the unit vector parallel to the
gradient of $\cF$. Expanding
$\cF(\vect{x}')\approx\cF(\vect{x})+\nabla\cF\cdot\dif\vect{x}$ and retaining
only the term linear in $\dif\vect{x}$, we reduce the above inequalities to
\[
F_\mathrm{c}-|\nabla\cF(\vect{x})|\,\dif x < \cF(\vect{x}) < F_\mathrm{c}\;,
\]
and, using arguments similar to those that lead to Eq.~(\ref{prob_over}) and
further, we obtain the probability of a crossing (either upward or downward)
of the contour $\cF(\vect{x})=F_\mathrm{c}$ in two dimensions:
\[
p_\mathrm{c}^{(2)}=\int_{-\infty}^\infty \int_{-\infty}^\infty
 f(F_\mathrm{c},\vect{V})\,|\vect{V}|\,\dif V_x \, \dif V_y \;
\]
where $f(F_\mathrm{c},\vect{V})$ is the joint probability distribution of
$\cF$ and its gradient. This quantity can be identified with the mean length
of the $F_\mathrm{c}$ contour per unit area,
\[
N_2=p_\mathrm{c}^{(2)}\;.
\]

In both one and two dimensions, the probability of crossing a contour is given
by the mean value of the magnitude of the gradient of the random field at that
contour. As discussed by Ryden (\cite{R88}), for an isotropic random field
$\cF$, the mean number of crossings per unit length along a line, $N_1$, and
the mean length of a contour per unit area, $N_2$, are related,
\[
N_2=\frac{\pi}{2}N_1\;,
\]
and so statistical properties of contours in the plane can be conveniently
studied using one-dimensional cuts, which is technically easier.

\subsubsection{The separation of canals and parameters of ISM turbulence}
Shukurov \& Berkhuijsen (\cite{Shukurov03}) applied a one-dimensional version
of this theory to canals observed at $\lambda20.5\cm$ in the direction to the
Andromeda nebula (around $l=120^\circ,b=-20^\circ$). Their estimate of the
mean separation of canals is $5'$ which results in a estimate of the Taylor
microscale of RM (or, more precisely, the Faraday depth) of
$l_\mathrm{T}\simeq2'$. For the presumed mean distance within the
Milky Way's non-thermal disc in that direction, 1\,kpc, the corresponding linear
scale is 0.6\,pc. If this scale can be identified with the Taylor microscale
of interstellar turbulence (which may or may not be the case!), the effective
value of the Reynolds number follows as $\mathrm{Re}\simeq(l_0/l_\mathrm{T})^2
\simeq10^5$ for the energy-range scale of $l_0=100\p$. This value of Re is many
orders of magnitude smaller than those based on the molecular viscosity. If
this estimate can be confirmed by more comprehensive and deeper analyses, this
would imply that the dissipation range of the interstellar MHD turbulence is
controlled by relatively large viscosities, perhaps associated
with kinetic plasma effects. The latter are often described as anomalous viscosity
or resistivity; useful recent results were obtained in the context of galaxy
clusters by Schekochihin \etal\ (\cite{S05}).

\subsection{Faraday screens: the mean separation of canals and shocks}
\label{subsec:shocks}
If canals are caused by rotation measure discontinuities in a Faraday
screen, the mean distance between them will be related to the
distribution of discontinuities in the ISM. Most plausibly these are
shock fronts. (Another possible source are tangential discontinuities
but it is difficult to identify a physical reason for their widespread
occurrence.)

A canal forms where the background Faraday depth $F$ changes by
$|\DF|=(n+1/2)\pi$ across a shock front. If the magnetic field is frozen into
the gas and the gas density increases by a factor $\er$ then so will the field
strength (we neglect a factor $\sim 1/\sqrt{2}$ due to the relative alignment
of the shock and field). Therefore
\begin{equation}
\label{eq:epsilon}
\er\simeq \sqrt{\left|\frac{\DF}{F_0}\right|+1}.
\end{equation}
The increase in gas density depends on the Mach number of the shock $\M$ as
(e.g. Landau \& Lifshitz~\cite{Landau60})
\begin{equation}
\label{eq:mach}
\er=\frac{(\gamma + 1) \M^2}{(\gamma - 1) \M^2 + 2}\simeq4\frac{\M^2}{\M^2+3}
\end{equation}
where we have used $\gamma\simeq 5/3$ for the polytropic index of the
gas. Combining Eqs.~(\ref{eq:epsilon}) and (\ref{eq:mach}), we obtain
the minimum Mach number necessary to produce a canal,
\begin{equation}
\label{eq:MRM}
\M_*\simeq \sqrt{\frac{3\er}{4-\er}}
        = \left[\frac{4}{3(|\DF/F_0|+1)^{1/2}} - \frac{1}{3}\right]^{-1/2}, \hspace{5mm} |\DF|=\frac{\pi}{2}.
\end{equation}
As follows from Sect.~\ref{subsec:screen}, stronger shocks can also produce canals.

The dominant source of the primary shocks are supernovae. When these shocks
encounter gas clouds they will be reflected, generating a population of
secondary shocks. Following Bykov \& Toptygin~(\cite{Bykov87}), the probability
density function of both primary and secondary shocks is given by
\begin{equation}
\label{eq:Pshock}
G(\M) \simeq G_0\left[\frac{1}{\M^{\alpha+1}}
        + \frac{3C(\alpha)f_\mathrm{cl}}{(\M-1)^4} \right],
\end{equation}
where $G_0=S\frac{4}{3}\pi r_0^3$, $S$ is the supernova rate per
unit volume, $r_0$ is the maximum radius of a supernova (primary) shock,
and $\alpha$ characterizes the dependence of the shock radius on the Mach
number, $r(\M)=r_0\M^{-\alpha/3}$, $f_\mathrm{cl}$ is the volume filling factor of
diffuse clouds and $C(\alpha)$ is a constant. For the Sedov solution,
$\alpha=2$ and for a three-phase ISM, $\alpha=4.5$; $C(\alpha)=2.3\times 10^{-2}$
for $\alpha=2$ and $4.1\times 10^{-3}$ for $\alpha=4.5$. The mean number of shocks
of a strength $\ge\M$ crossing a position in the ISM per unit time is
\begin{equation}
\label{eq:Fshock}
H(\M)=\int_\M^\infty G(\M)\,\dif\M=G_0\left[\frac{1}{\alpha\M^{\alpha}}
        + \frac{C(\alpha)f_\mathrm{cl}}{(\M-1)^3} \right],
\end{equation}
and the mean separation of shocks with a strength $\ge\M$ is
\begin{equation}
\label{eq:Lshock}
L(\M)\simeq\frac{c_s}{H(\M)},
\end{equation}
with $c_s$ the sound speed.
Substituting Eqs.~(\ref{eq:MRM}) and (\ref{eq:Fshock}) into
Eq.~(\ref{eq:Lshock}), we find the mean separation of shocks in three
dimensions:
\begin{eqnarray}
\label{eq:Lshock3D}
L&\simeq& 9\p \left(\frac{c_s}{10\kms}\right)
        \left(\frac{\nu_\mathrm{SN}}{0.02\yr^{-1}}\right)^{-1}
        \left(\frac{R}{15\kpc}\right)^{2}
        \left(\frac{h}{50\p}\right)
        \left(\frac{r_0}{100\p}\right)^{-3}\nonumber\\
         &&\mbox{}\times\left[\frac{1}{\alpha\M_*^{\alpha}}
         + \frac{C(\alpha)f_\mathrm{cl}}{(\M_*-1)^3} \right]^{-1},
\end{eqnarray}
where $\nu_\mathrm{SN}$ is the Galactic supernova rate, $R$ and $h$ are
the radius and scale height of the star-forming disc respectively;
the term in the square brackets is close 0.2 for $\M=1.2$ and $\alpha=4.5$.
Projecting onto the sky plane changes $\Lc$ by a factor
$\xi=\pi^{-1}\int_0^\pi\sin^2\theta\,\dif\theta=1/2$. If the shocks fill
a region of a depth $d$, the separation of the shocks in the plane of
the sky is further reduced by a factor $d/L$. The angular separation
$\Lc$ at an average distance $d/2$ to the shocks follows as
\begin{equation}
\label{eq:Lshock2D}
\Lc=\frac{L^2}{d^2}
\end{equation}

As an example, let us use the observations by Haverkorn \etal\
(\cite{Haverkorn03}) as inputs to Eq.~(\ref{eq:Lshock2D}) and compare
the observed separation of canals with the predicted separation of
shocks. In this field the average Faraday depth is $F\simeq \pi\,
\mathrm{radians}$ and the most abundant canals are produced when $|\DF|=
\pi/2$, since weaker shocks are the most abundant according to
Eq.~(\ref{eq:Pshock}). Equation~(\ref{eq:MRM}) then gives $\M_*\simeq
1.2$. The maximum depth of the layer is estimated as $d\simeq 600\p$.
With the parameter values used to normalize Eq.~(\ref{eq:Lshock3D}) and
$f_\mathrm{cl}=0.25$, $\alpha=4.5$, we obtain a mean separation of canals of
$L\simeq 40\p$, or $\Lc\simeq 15^{\prime}$. The latter value is
reasonably close to a
by-eye estimate of the average separation of canals in Fig.~3 of
Haverkorn \etal\ (\cite{Haverkorn03}) of $45^{\prime}$. Thus, the canals
observed by Haverkorn \etal\ (\cite{Haverkorn03}) are compatible with a
system of shocks producing discontinuities in a Faraday screen. The
canals in these observations appear to be straighter (or less twisting)
than those observed at shorter wavelengths by, e.g., Uyan{\i}ker \etal\
(\cite{Uyaniker98a}) and Gaensler\etal\ (\cite{Gaensler01}). This might
be expected if the straighter canals arise from shocks in the diffuse
ISM and the twisting canals have a different origin such as differential
Faraday rotation. The Mach number of the shocks required to produce
$|\DF|=\pi/2$ at metre wavelengths is quite low, so that they can be
inefficient in accelerating cosmic rays. Then the expected increase in total
synchrotron emissivity at a shock is just $\M_*^2\simeq 1.5$. Such features can be difficult to
detect in total intensity radio maps, especially if their area covering
factor is high.

\section{Prospects}
At first sight, the chaotic appearance of radio polarization maps of the
diffuse Milky Way emission and the difficulty of connecting standard
statistical tools, such as power spectra, to ISM physics, makes interpretation
of the data a daunting task. Moreover, changes in the polarization pattern
with wavelength depend on random quantities, such as the thermal electron
density and magnetic field fluctuations. Analysis of the ubiquitous
depolarization canals provides a practical starting point from which we can
develop the methods required to make full use of the polarization
observations. As we have shown here, the origin of canals can be determined
from observed quantities --- by examining how $Q$ and $U$ vary across a canal
--- and their statistical measures are related to the parameters of
interstellar turbulence. The extension of methods used to study canals to the
wider analysis of polarized radio emission will lead to better techniques for
separating Milky Way foregrounds from polarized cosmological signals.

\section*{Acknowledgements}
This work was supported by the Leverhulme Trust under research grant
F/00~125/N. We thank Wolfgang Reich for providing Figure 1 and N.~Makarenko
for useful discussions on contour statistics.



\begin{thebibliography}{}

\bibitem[1986]{B86} Barden J.~M., Bond J.~R., Kaiser N. \& Szalay A.~S., 1986,
ApJ, 304, 15

\bibitem[1999]{Beck99} Beck R., 1999, in {\it ``Galactic
Foreground Polarization''}, ed.\  E.~M.~Berkhuijsen (MPIfR: Bonn), p.\ 3

\bibitem[1966]{Burn66} Burn B.\ J., 1966, MNRAS, 133, 67

\bibitem[1987]{Bykov87} Bykov A.\ M.\ \& Toptygin N., 1987, Ap\&SS, 138, 341

\bibitem[1987]{CB87} Coles P.\ \& Barrow J.~D., 1987, MNRAS, 228, 407

\bibitem[1999]{Duncan99} Duncan A.\ R., Reich P., Reich W.\ \& F\"urst E., 1999,
A\&A, 350, 447

\bibitem[1989]{E89} Eilek J.~A., 1989, AJ, 98, 244

\bibitem[2001]{Gaensler01} Gaensler B.\ M., Dickey J.\ M., McClure-Griffiths N.\
M., Green A.\ J., Wieringa M.\ H.\ \& Haynes R.\ F., 2001, ApJ, 549, 959

\bibitem[1999]{Gray99} Gray A.\ D., Landecker T.\ L., Dewdney P.\ E., Taylor A.\
R., Willis A.\ G.\ \& Normandeau M., 1999, ApJ, 514, 221

\bibitem[2004]{Heitsch04} Haverkorn M.\ \& Heitsch F., 2004, A\&A, 421, 1011

\bibitem[2000]{Haverkorn00} Haverkorn M., Katgert P.\ \& de Bruyn A.\ G., 2000,
A\&A, 356, L13

\bibitem[2003]{Haverkorn03} Haverkorn M., Katgert P.\ \& de Bruyn A.\ G., 2003,
A\&A, 403, 1031

\bibitem[2004]{Haverkorn04} Haverkorn M., Katgert P.\ \& de Bruyn A.\ G., 2004,
A\&A, 427, 549

\bibitem[1960]{Landau60} Landau L.\ D.\ \& Lifshitz E.\ M., 1960, \textit{``Fluid
Mechanics''} (Pergamon Press: Oxford)

\bibitem[1957]{Longuet57} Longuet-Higgins M.\ S., 1957, Phil. Trans. R. Soc. London, Ser. A, 249, 321

\bibitem[1984]{P84} Peebles P.~J.~E., 1984, ApJ, 277, 470

\bibitem[2004]{Reich04} Reich, W., F\"urst, E., Reich, P., Uyan{\i}ker,  B.,
Wielebinski, R.\ \& Wolleben, M., 2004, in \textit{``The Magnetized Interstellar
Medium''}, ed.\ B.\ Uyan{\i}ker, W.\ Reich \& R.\ Wielebinski
(Katlenburg-Lindau: Copernicus), p.~45

\bibitem[1988]{R88} Ryden B.\ S., 1988, ApJ, 333, L41

\bibitem[1989]{R89} Ryden B.\ S., Melott A.\ L., Craig D.\ A., Gott R.,
Weinberg D.~H., Scherrer R.~J., Bhavsar S.~P.\ \& Miller J.~M., 1989, ApJ,
340, 647

\bibitem[2005]{S05} Schekochihin A.~A., Cowley S.~C., Kulsrud R.~M., Hammett G.~W.\
\& Sharma P., 2005, ApJ, 629, 139

\bibitem[2003]{Shukurov03} Shukurov A.\ \& Berkhuijsen E.\ M., 2003, MNRAS, 342,
496 (Erratum: 2003, MNRAS, 345, 1392)

\bibitem[1998]{Sokoloff98} Sokoloff D.\ D., Bykov A.\ A., Shukurov A.,
Berkhuijsen E.\ M., Beck R.\ \& Poezd A.\ D., 1998, MNRAS, 299, 189 (Erratum: 1999,
MNRAS, 303, 207)

\bibitem[1966]{Sveshnikov66} Sveshnikov A.~A., 1966, {\it ``Applied Methods in the Theory of
Random Functions''} (Pergamon Press: Oxford)

\bibitem[1972]{TL72} Tennekes H., Lumley J.~L., 1982, \textit{``A First
Course in Turbulence''} (MIT Press: Cambridge, Mass)

\bibitem[1998a]{Uyaniker98b} Uyan\i ker B., F\"urst E., Reich W., Reich P.\ \&
Wielebinski R., 1998a, A\&AS, 132, 401

\bibitem[1998b]{Uyaniker98a} Uyan{\i}ker B., F\"urst E., Reich W., Reich P.\ \&
Wielebinski R., 1998b, A\&AS, 138, 31

\bibitem[1983]{Vanmarcke83} Vanmarcke E., 1983, \textit{``Random Fields:
Analysis and Synthesis''} (MIT Press: Cambridge, Mass.)

\bibitem[1993]{Wieringa93} Wieringa M.\ H., de Bruyn A.\ G., Jansen D., Brouw W.\
N.\ \& Katgert P., 1993, A\&A, 268, 215

\bibitem[2006]{Wolleben06} Wolleben M., Landecker T.\ L., Reich W.\ \&
Wielebinski R., 2006, A\&A, 448, 411

\end{thebibliography}
\end{document}